# A Novel Loss Function Incorporating Imaging Acquisition Physics for PET Attenuation Map Generation using Deep Learning

Luyao Shi[1], John A. Onofrey[1], Enette Mae Revilla[1], Takuya Toyonaga[1], David Menard[2], Joseph Ankrah[2], Richard E. Carson[1], Chi Liu[1] and Yihuan Lu[1(✉)]

[1] Yale University, New Haven, CT 06520, USA
`Yihuan.lu@yale.edu`
[2] Yale New Haven Hospital, New Haven, CT 06520, USA

**Abstract.** In PET/CT imaging, CT is used for PET attenuation correction (AC). Mismatch between CT and PET due to patient body motion results in AC artifacts. In addition, artifact caused by metal, beam-hardening and count-starving in CT itself also introduces inaccurate AC for PET. Maximum likelihood reconstruction of activity and attenuation (MLAA) was proposed to solve those issues by simultaneously reconstructing tracer activity (λ-MLAA) and attenuation map (μ-MLAA) based on the PET raw data only. However, μ-MLAA suffers from high noise and λ-MLAA suffers from large bias as compared to the reconstruction using the CT-based attenuation map (μ-CT). Recently, a convolutional neural network (CNN) was applied to predict the CT attenuation map (μ-CNN) from λ-MLAA and μ-MLAA, in which an image-domain loss (IM-loss) function between the μ-CNN and the ground truth μ-CT was used. However, IM-loss does not directly measure the AC errors according to the PET attenuation physics, where the line-integral projection of the attenuation map (μ) along the path of the two annihilation events, instead of the μ itself, is used for AC. Therefore, a network trained with the IM-loss may yield suboptimal performance in the μ generation. Here, we propose a novel line-integral projection loss (LIP-loss) function that incorporates the PET attenuation physics for μ generation. Eighty training and twenty testing datasets of whole-body [18]F-FDG PET and paired ground truth μ-CT were used. Quantitative evaluations showed that the model trained with the additional LIP-loss was able to significantly outperform the model trained solely based on the IM-loss function.

## 1 Introduction

Positron Emission Tomography (PET) is used to assess physiological or pathological processes, e.g., cancer staging, via the use of specific tracers. Those assessments rely on *in vivo* radiotracer quantification based on PET images, which require accurate attenuation correction (AC). CT-based AC is commonly used for PET/CT studies. However, mismatch between CT and PET due to patient body motion [1] results in AC artifacts and thus inaccurate PET quantification. In addition, artifact caused by metal, beam-hardening and count-starving in CT itself also introduces inaccurate AC for PET. Therefore, a method that can generate an accurate attenuation map (µ) is critical for accurate radiotracer quantification in PET.

The maximum likelihood reconstruction of activity and attenuation (MLAA) algorithm [2] was proposed to simultaneously reconstruct tracer activity (λ-MLAA) and artifact-free attenuation map (µ-MLAA) based on the time-of-flight (TOF) PET raw data only. However, mainly due to the limited TOF timing resolution, µ-MLAA suffers from high noise and λ-MLAA suffers from large bias [3] as compared to the standard maximum likelihood expectation maximization reconstruction using the CT-based attenuation map (µ-CT). Recently, Hwang et al. [4] proposed to use a convolutional neural network (CNN) to predict the CT attenuation map (µ-CNN) from λ-MLAA and µ-MLAA. Similar to other image-to-image translation work [5, 6], Hwang et al. [4] used the L1-norm between the µ-CNN and the ground truth µ-CT in the image domain as a loss function (IM-loss). However, this loss choice does not directly measure the AC errors according to the PET attenuation physics, where the line-integral of the µ along the path of the two annihilation events, instead of µ itself, is used for AC. Therefore, a CNN trained using the IM-loss function may yield suboptimal performance in the µ generation task. In this study, we propose a novel loss function that uses the line-integral projection (LIP-loss) of µ as the loss, in addition to the IM-loss, for µ generation. Our hypothesis is that the additional LIP-loss will provide a stronger constraint than the IM-loss alone for µ generation, and therefore a more accurate µ can be generated.

We evaluated our method on real whole-body PET/CT datasets. Experimental results demonstrate that incorporating the proposed LIP-loss in the network training significantly improved the accuracy of the predicted µ and yielded more accurate quantification in the final attenuation corrected PET images compared with the conventional training strategy using only the IM-loss.

## 2   Datasets

For this study, 220 whole-body, i.e., skull to feet, $^{18}$F-FDG PET/CT scan data of patients were acquired using a Siemens Biograph mCT 40 scanner. Based on careful human-observer examination, i.e., visual comparison between λ-MLAA/µ-MLAA and CT, 100 scans with minimal mismatch, i.e., body motion free, between the ground truth µ-CT and PET, were selected for training (N=80) and testing (N=20). The CTs for the 100 selected scans were also without artifacts. All scans were performed ~60 min after intravenous injection of ~10 mCi $^{18}$F-FDG. The entire body of each patient with arm-down position was scanned using the continuous bed motion protocol for ~20 min. For MLAA, we used the same implementation as in [3] with 3 iterations by 21 subsets. Both λ-MLAA and µ-MLAA were originally reconstructed using 2 mm voxel size followed by 5 mm Gaussian post-smoothing, and were further down-sampled to 4 mm. CT attenuation maps were generated using the Siemens e7 toolkit and down-sampled to 4 mm voxel in width to save GPU memory in the later network training.

3## 3 Methods

### 3.1 Line-integral Projection Loss (LIP-loss) Function

The LIP-loss measures the line-integral difference between the image patch $X$ of the predicted µ and the ground-truth patch $Y$ of µ-CT in the projection domain:

$$L_{LIP}(X,Y) = \frac{1}{N_\Bbbk}\sum_{k\in\Bbbk}\frac{\sum_{i=1}^{N_k}([P_kX]_i-[P_kY]_i)^2}{N_k} \tag{1}$$

where $\Bbbk$ is the set of line-integral projection (LIP) angles, $k$ is the index for the projection angles, $N_\Bbbk$ is the total number of angles in $\Bbbk$, $P_k$ is the LIP operator on the image $X$ and $Y$ at the $k$-th angle, $i$ is the pixel index in the projection domain and $N_k$ is the total number of pixels in the LIP $P_kX$ and $P_kY$. Set $\Bbbk$ was designed such that $N_\Bbbk$ angles were uniformly sampled over 180 degrees, i.e., $\Bbbk = \{0°, 45°, 90°, 135°\}$ in the case of $N_\Bbbk=4$. In our implementation, we rotated the images (using bilinear interpolation) and perform LIP at a single angle, instead of performing LIP at different angles. Note that the LIP-loss can be easily back-propagated to update the weights in the network, since the LIP operator $P_k$ is a linear operation so that the loss function is differentiable.

In terms of loss function, the conventional IM-loss is constructed as

$$L_{IM}(X,Y) = L_{L1}(X,Y) + \lambda_1 L_{GDL}(X,Y), \tag{2}$$

where $L_{L1}$ is an L1-norm loss, which was reported [5, 6] to better preserve anatomical structures than an L2-norm loss. $L_{GDL}$ is an image gradient difference loss defined as:

$$L_{GDL}(X,Y) = \bigl||\nabla X_x|-|\nabla Y_x|\bigr|^2 + \bigl||\nabla X_y|-|\nabla Y_y|\bigr|^2 + \bigl||\nabla X_z|-|\nabla Z_z|\bigr|^2, \tag{3}$$

where $\nabla$ is the gradient operator. $L_{GDL}$ is used to further discourage image blurring [7].

To enforce the additional similarity in the projection domain between the predicted and ground-truth µ-CT, the proposed LIP-loss ($L_{LIP}$) was added to the IM-loss as:

$$L_{TOTAL}(X,Y) = L_{L1}(X,Y) + \lambda_1 L_{GDL}(X,Y) + \lambda_2 L_{LIP}(X,Y), \tag{4}$$

where $\lambda_1$ and $\lambda_2$ are the weights for the $L_{GDL}$ and $L_{LIP}$ terms, respectively. The proposed framework for the training phase is illustrated in Fig.1. In this paper, we refer the proposed method, i.e., training using $L_{TOTAL}$, as **l**ine-**i**ntegral **p**rojection enforced **d**eep **l**earning method (LIPDL), and the conventional method, i.e., training using $L_{IM}$, as deep learning method (DL).

### 3.2 Network Architectures

In this work, we used a modified version of the fully-convolutional U-net architecture [8] for predicting the attenuation map from λ-MLAA and µ-MLAA. The network operates on 3D patches and uses 3×3×3 convolution kernels. Different from the original U-net, where 2×2×2 max pooling operations are used at the end of each stage, we reduced the resolution by using convolution operations with 2×2×2 kernels and stride 2 [9]. In addition, symmetric padding was applied to the input image (and the feature maps in later layers) prior to the convolution operations to avoid reducing the image (or feature map) sizes due to the convolution. This allows the network's output layer to have the same size as the input layer. Batch normalization was applied after each convolutional layer and before the ReLU. Dropout with a rate of 0.15 was applied to the bottleneck layer of the U-net in the training phase to prevent overfitting, however, was removed in the testing phase.

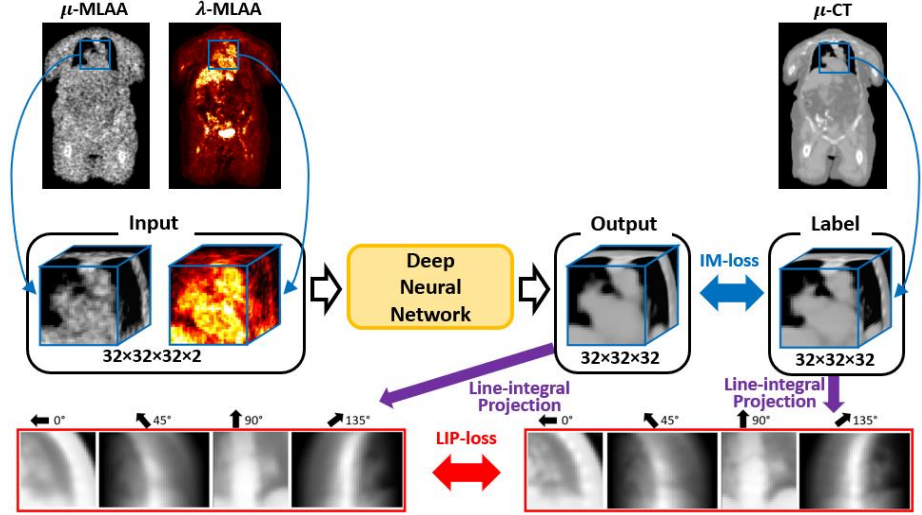

**Fig. 1.** Proposed framework (training phase). Both image domain loss (IM-loss) and line-integral projection loss (LIP-loss) are used to update the deep neural network.

### 3.3 Image preprocessing

Image normalization is a key pre-processing step for deep learning algorithms [10]. Unlike CT images, in which the image intensity (Hounsfield Unit, HU) represents attenuation as relative to water and the intensity range is consistent across all the patients, the PET image intensity represents the tracer uptake level. The use of standardized uptake value (SUV) in PET helps to normalize the tracer injection dose and patient weight, however, the biological uptake range is intrinsically broad for FDG PET, e.g., the contrast between brain and muscle can be 10:1, and even 100:1 between bladder and muscle tissue. Additional image normalization is needed in order to obtain more stable results. In this study, the λ-MLAA images were normalized using $\lambda_{norm} = \tanh(\lambda/\sigma)$ before training and testing, where $\lambda$ and $\lambda_{norm}$ are the λ-MLAA images (in SUV) before and after normalization; $\sigma$ is a parameter controlling the range of the active gradient zone of the hyperbolic tangent (tanh) function, which was empirically set to 5 to ensure that the organs of interests, i.e., except bladder, are in the active gradient zone. The μ-CT and μ-MLAA images were normalized by 0.15 cm$^{-1}$, which corresponds to skull bone attenuation coefficient at 511keV, to match the value range of $\lambda_{norm}$. The normalized λ-MLAA and μ-MLAA were concatenated as a multi-channel image and used as the input to the deep neural networks for training and testing.

### 3.4 Evaluation

In this study, 80 subjects were included in training, and 20 subjects were used for evaluation. The predicted attenuation maps from the proposed LIPDL method (μ-LIPDL) were compared with those trained using only the image domain loss $L_{IM}$ (μ-DL) and μ-MLAA, using the μ-CT as the reference.

The quality of the predicted attenuation maps was evaluated regarding the normalized mean absolute error (NMAE), the mean squared error (MSE), peak signal-to-noise ratio (PSNR), structural similarity index (SSIM), line-integral normalized mean absolute error (LINMAE) and line-integral mean squared error (LIMSE). The NMAE was defined as $\text{NMAE} = (\sum_{x,y,z}|X(x,y,z) - Y(x,y,z)|)/(N(\max(Y) - \min(Y)))$, where X and Y present the predicted attenuation map and the reference CT-attenuation map, $max$ and $min$ operators calculate the maximum and minimum intensities of the reference image. $N$ is the total number of voxels. The LINMAE and LIMSE measure the error in line-integral projection domain. LINMAE was defined as: $\text{LINMAE} = [\sum_{k\in\mathbb{k}}\sum_{i=1}^{N_k}|[P_kX]_i - [P_kY]_i|/(N_k(\max(P_kY) - \min(P_kY)))]/N_{\mathbb{k}}$. The definition of LIMSE can be found in Eq.1.

For each patient, 4 PET reconstructions, using the ordered subset expectation maximization (3 iterations by 21 subsets, 5 mm Gaussian smoothing) algorithm, were performed using μ-LIPDL, μ-DL, μ-MLAA, and the ground-truth μ-CT as the attenuation map, respectively. All the attenuation maps were resliced to 2 mm voxel in width prior to the PET reconstructions. 2 mm voxel in width was used in PET reconstruction. Using the PET reconstructed with μ-CT as the reference, NMAE and MSE were computed for μ-LIPDL, μ-DL and μ-MLAA, respectively, on the entire body as well as 5 anatomical regions: head, neck to chest, abdomen, pelvis and legs, which correspond to the 0%-10%, 10%-30%, 30%-40%, 40%-50% and 50%-100% segments of each patient.

## 4 Experimental Results

For both LIPDL and DL methods, we trained the networks for 80 epochs, respectively. In each epoch, 40,000 32×32×32 patches were randomly sampled from the training data and batch size of 16 was used for updating the network. The networks were trained with the Adam optimizer. An initial learning rate of $10^{-3}$ was used, which was decayed by a factor of 0.99 after each epoch. $\lambda_1$ and $\lambda_2$ were set to 1 and 0.02, respectively. In the testing phase, to reduce the stitching artifacts caused by overlapping small image patches, we used relatively a large patch size of 200×200×32 and stride size of 200×200×16 (i.e., no striding in the first 2 dimensions). We implemented our framework using Tensorflow. The training takes about 40 hours on an NVIDIA GTX 1080 Ti GPU.

## 4.1 Attenuation Map Evaluation

Fig.2 shows one example of different attenuations maps. Qualitatively, both μ-DL and the proposed μ-LIPDL yielded much less noisy attenuation maps as compared to the μ-MLAA, and visually, both μ-DL and μ-LIPDL are very similar to the μ-CT. However, μ-LIPDL showed more consistent intestine cavity area (yellow arrow) than the μ-DL as compared to the μ-CT. Note that the CT reconstruction artifacts in μ-CT (red arrow) were also removed in both μ-DL and μ-LIPDL.

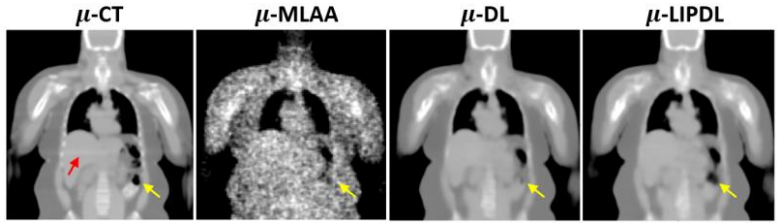

**Fig. 2.** Visual comparison for the original μ-CT, the μ-MLAA, the predicted μ-DL and μ-LIPDL. Improved intestine cavity area (yellow arrow) prediction can be seen in the proposed μ-LIPDL, as compared to the standard μ-DL. Red arrow in the μ-CT points at the CT reconstruction artifacts (mild). The same gray-scale window is used for all images.

**Table 1.** Evaluation of the attenuation maps generated by three methods over 20 subjects using the image domain quality metrics and the line integral projection (LIP) domain quality metrics.

|  | Metric | μ-MLAA | μ-DL | μ-LIPDL |
|---|---|---|---|---|
| Image domain quality metrics | NMAE | 6.29%±1.16% | 1.66%±0.25% | **1.61%±0.24%**\* |
|  | MSE | 3.26E-4±6.22E-5 | 3.20E-5±9.24E-6 | 3.23E-5±9.66E-6 |
|  | PSNR | 28.1±1.06 | 38.2±1.02 | 38.2±1.06 |
|  | SSIM | 93.1%±0.007% | 99.8%±0.0009% | 99.8%±0.0008% |
| LIP domain quality metrics | LINMAE | 7.95%±1.10% | 1.29%±0.24% | **1.08%±0.19%**\*\* |
|  | LIMSE | 0.769±0.144 | 0.024±0.010 | **0.018±0.008**\*\* |
| mean ± STD,*indicates μ-LIPDL and μ-DL's difference is significant (\*: $p<10^{-5}$, \*\*: $p<10^{-8}$) | | | | |

Table 1 quantitatively shows that both μ-DL and μ-LIPDL yielded statistically significant superior performance than the μ-MLAA over all evaluation metrics (maximum $p<10^{-13}$). A paired t-test was used to evaluate statistical significance cross subjects (N=20). Interestingly, no significant difference was found between μ-DL and μ-LIPDL when the conventional image domain quality metrics, i.e., MSE, PSNR and SSIM, were used. Only 3% reduced error from μ-DL to μ-LIPDL was observed for NMAE although it was statistically significant. In contrast, when the line-integral domain quality metrics were used, large improvements in μ-LIPDL were found as compared to μ-DL. Specifically, μ-LIPDL yielded a 16.3% reduction in LIMAE and 33.3% reduction in LIMSE as compared to the μ-DL. We note that for the PET attenuation correction (AC) task, the line integral of μ is used, therefore, an attenuation map yielding lower line-integral error will provide superior performance in the AC task than one with larger line-integral error. These results suggest that training with only the conventional image domain loss might produce suboptimal results, since the image domain loss cannot distinguish a better attenuation map, i.e., µ-LIPDL, than a suboptimal attenuation map, i.e., µ-DL, for the AC task.

### 4.2 Attenuation Correction Performance in PET Reconstruction

PET images corrected by µ-LIPDL and µ-DL yielded significantly lower NMAE and MSE than those corrected by µ-MLAA, respectively (maximum $p<10^{-3}$). Reconstructed PET images corrected by the µ-LIPDL yielded statistically significant lower NMAE and MSE than those corrected by µ-DL. As shown in Table 2, under both metrics, the LIPDL-obtained results with substantially and significantly smaller errors than DL on all the 5 body regions as well as on the whole body.

**Table 2.** Using PET images corrected by the µ-CT as the reference, the NMAE and MSE of the PET images corrected by µ-MLAA, µ-LIPDL and µ-DL, respectively, were shown. Evaluations were performed on 5 different anatomical regions as well as on the whole body.

| | AC method | Head | Neck/Chest | Abdomen | Pelvis | Legs | Whole-body |
|---|---|---|---|---|---|---|---|
| NMAE | µ-MLAA | 21.4% | 7.8% | 9.4% | 9.2% | 8.5% | 11.26% |
| | µ-DL | 3.5% | 4.2% | 4.8% | 4.4% | 3.6% | 4.1% |
| | µ-LIPDL | **3.2%** * | **3.7%** ** | **4.1%** * | **3.6%** *** | **3.2%** ** | **3.6%** *** |
| MSE | µ-MLAA | 1.3E-01 | 7.4E-03 | 1.2E-02 | 1.4E-02 | 3.3E-03 | 3.4E-02 |
| | µ-DL | 2.7E-02 | 3.5E-03 | 4.1E-03 | 7.7E-03 | 9.2E-04 | 8.7E-03 |
| | µ-LIPDL | **1.9E-02*** | **2.7E-03*** | **3.1E-03*** | **4.8E-03*** | **7.4E-04*** | **6.0E-03**** |

\* indicates µ-LIPDL and µ-DL's difference is significant (\*:$p<10^{-2}$, \*\*:$p<10^{-4}$, \*\*\*:$p<10^{-6}$)

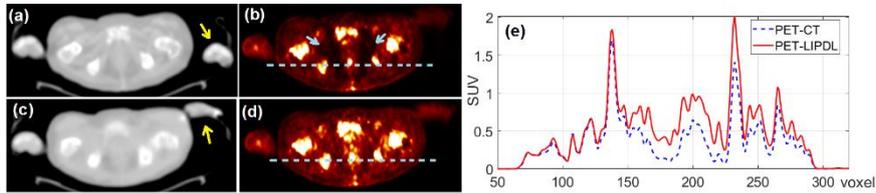

**Fig. 3.** (a) Coronal µ-CT, (b) PET corrected by µ-CT, (c) µ-LIPDL, (d) PET corrected by µ-LIPDL and (e) line profile measured on PET images corrected by µ-CT and µ-LIPDL.

### 4.3 Clinical Impact

Clinically, patient motion introduced mismatch between PET and CT that not only can generate AC inaccuracy in the PET reconstruction, but also can result in scatter correction (SC) inaccuracy. Fig.3 shows a case that patient right arm moved substantially between CT (a) and PET (b). Such mismatch caused erroneous SC, which is compounded with the AC artifact in the PET (dark area, arrows in (b)). µ-LIPDL (c) removed such mismatch and yielded stable SC in the PET reconstruction (d). Fig.3 (e) shows the line profile (dashed line in (b) and (d)) comparison between the PET corrected by µ-CT and µ-LIPDL. High uptakes in the PET indicate bone metastasis.

## 5 Summary

We have proposed a novel line integral loss function which incorporates imaging acquisition physics for PET attenuation map generation using deep learning. We showed that by enforcing the image projection domain consistency while training the deep neural networks, the generated attenuation maps perform significantly better for the task of PET attenuation correction, compared with conventional training that focuses solely on image domain consistency. In this study we used a modified version of U-net to demonstrate the effectiveness of the proposed training strategy, although the proposed method can be applied with any other neural networks. At the point of writing this paper, we empirically set the weight of the proposed line integral loss to 0.02 and obtained substantial improvement, we anticipate that fine tuning this parameter in the future could further improve the results. Furthermore, the proposed method is not only applicable to PET-CT image synthesis, but also to MRI-CT synthesis (for PET/MRI systems) for the purpose of generating attenuation maps.